# Dynamic Knowledge Selector and Evaluator for recommendation with Knowledge Graph

**Feng Xia**

Feng.xia.cs@gmail.com

**Zhifei Hu**

knkw25@gmail.com

A B S T R A C T

In recent years recommendation systems typically employ the edge information provided by knowledge graphs combined with the advantages of high-order connectivity of graph networks in the recommendation field. However, this method is limited by the sparsity of labels, cannot learn the graph structure well, and a large number of noisy entities in the knowledge graph will affect the accuracy of the recommendation results. In order to alleviate the above problems, we propose a dynamic knowledge-selecting and evaluating method guided by collaborative signals to distill information in the knowledge graph. Specifically, we use a Chain Route Evaluator to evaluate the contributions of different neighborhoods for the recommendation task and employ a Knowledge Selector strategy to filter the less informative knowledge before evaluating. We conduct baseline model comparison and experimental ablation evaluations on three public datasets. The experiments demonstrate that our proposed model outperforms current state-of-the-art baseline models, and each module's effectiveness in our model is demonstrated through ablation experiments.

## 1. Introduction

The recommendation algorithm is a vital component in applications such as e-commerce[1-3], news[4-6] and short videos[7-8]. As part, it takes on the important task of outputting personalized information flow. Early recommendation algorithms[9-12] often employed the K-nearest neighbor method, such as item-based collaborative filtering algorithm[13-16] by comparing the similarity between items and recommend the most similar items to the historical click items of the user. The recommendation algorithm is based on the nearest neighbor search method, which benefits from strong interpretability, high performance, and simplicity, but it is not able to analyze user characteristics, item characteristics, and context characteristics models, and it performs poorly in recommendation scenarios with fewer user behavior data.

Introducing the knowledge graph[17-18] to the recommendation system[19-25] effectually reduces the cold start problem in the recommendation system. Because the rich semantic relationship information between entities in the knowledge graph provides a large amount of auxiliary information for the items (i.e., side information), it hence results in rising the richness of the recommended items and enhancing a recommendation algorithm based on knowledge graphs. Shi et al.[26] designed a meta-path-based random walk sampling method, generating multiple nodes starting from the target node, and consisting of multiple nodes such that the points constitute the meta-path. On this basis, a meta-path embedding learning algorithm (i.e., heterogeneous information network (HIN)) was proposed. Sun [27] proposed a loop-based network approach, which represents a modular algorithm (i.e., recurrent knowledge graph embedding (RKGE)) that compensates for the embedding learning algorithm. This methodology only cares about the semantic layer connectivity of nodes, but overlooks the path information between nodes. Wang [28] proposed a method for modeling path order dependencies (i.e., knowledge-aware recurrent network (KPRN)), which can model the complex relationship of paths connecting users and items and combine them through pooling operations on the importance of various paths. Wang [29] proposed a method based on the knowledge graph and attention machine, a system-based algorithm (i.e., knowledge graph attention network (KGAT)), two parts between users and items. The graph is then combined with the knowledge graph and the graph relationship, whereas users and items are displayed in the same graph space. By employing a network graph attention mechanism to extract high-level and semantic relationships between entities from mixed graphs, Wang [30] proposed a model based on the perceptual domain (the so-called knowledge graph attention network (KGCN)). According to the user's historical click on the item, it looks for neighboring nodes in the knowledge map, employs the graph attention mechanism to continuously collect, and then updates and disseminates information. Wang et al. [31] proposed an algorithm model for label smoothing (i.e., knowledge-aware graph neural networks with label smoothing regularization (KGNN-LS)). This model introduces a regular label smoothing term based on the KGCN to adjust the loss function constraints. Wang et al. [32] proposed a water wave network (RippleNet), which is cleverly borrowed from water wave transmission. The broadcasting idea, using the items that the user has clicked on in the history as the user's interest seeds, broadcasts the items knowledge map to the outer layer, combines the representation-based learning method with the path-based method, and achieve to propagate user preferences to discover potential interests of users.

The complexity algorithm of the above graph is based on the mechanism of aggregation-propagation-update, which is very complicated for calculation, and the importance assessment of each entity is not enough, and the information of each entity cannot be utilized. We will explain this in some detail in the motivation section.

Based on the above research background, this paper also aims to propose a dynamic knowledge selection and evaluator to extract information from knowledge graphs with the help of supervised information. First, all relationships and entities between each central entity and neighboring entities are taken into account as knowledge links. The knowledge selector is employed to select the information that represents the importance of this link, the evaluator is then utilized to score the importance of each entity, and finally, all entities are appropriately aggregated as context information of the current entity to increase the semantic representation. We conduct experiments on three open-source datasets. A large number of experimental results indicate the effectiveness of the proposed model. Meanwhile, we design a set of ablation tests for the internal analysis of the model. Our contributions and innovations are:

- A dynamic knowledge selection and evaluator is proposed to maximize the use of structural information in KG for knowledge distillation .

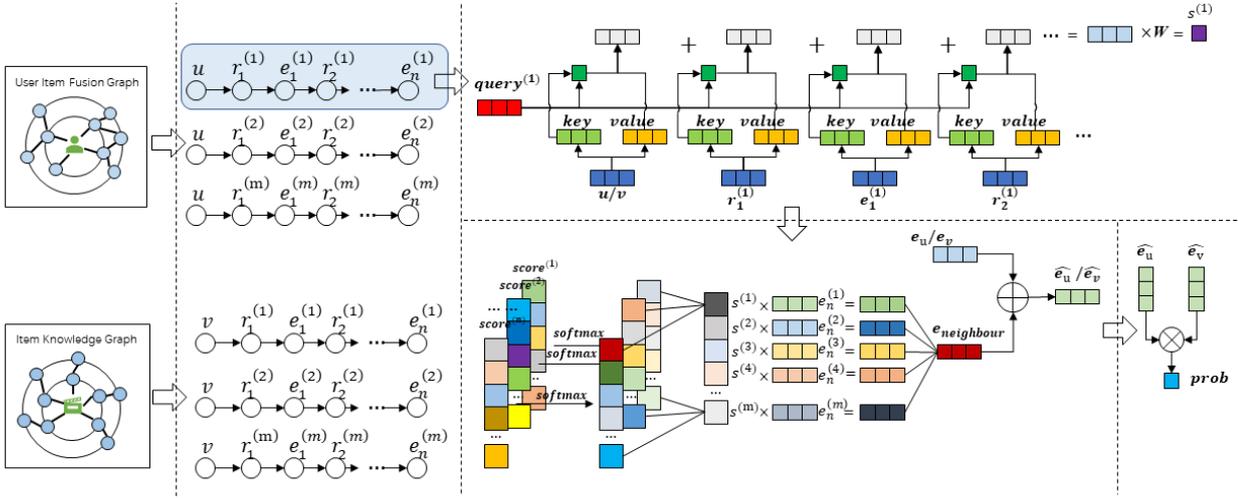

**Figure 1: The framework of the proposed DKSE. The framework is composed of three major modules: Knowledge Selector , Chain Route Evaluator and Prediction module.**

- A large number of experiments are conducted on three datasets to prove the superiority of the model compared to the current best model.

## 2. Motivation

This section revisits existing KG-based recommendation methods and further identifies the limitations originating from the information aggregation attention mechanism, clarifying the motivation of our work.

### 2.1. Revisiting KG-based recommendation methods

KG-based recommendation methods recursively propagate an embedding from its neighbors to refine the embedding. Moreover, such methods calculate the inner product of user and item representations to predict their matching score and employ specific attention mechanisms to discriminate the neighbors' importance. KG-based recommendation methods can be formulated as follows:

**Knowledge-aware Attention:** Information propagation between an entity and its neighbors can be achieved by computing the normalized score $\tilde{\pi}$ to control the decay factor on the edge, indicating how much information is propagated from the tail entity. The score $\pi_e$ is computed using function $f_s$, which is specifically designed for this purpose.

$$\tilde{\pi} = \frac{\exp(\pi_e)}{\sum_{e \in N(v)} \pi_e} \tag{1}$$

**Information Aggregation:** After specifying the varying importance of the neighbors, the entity and neighbor representations are aggregated using the aggregator $f_{agg}$.

**Information Propagation:** To explore the high-order connectivity information and collect the information propagated from the higher-order neighbors, the embedding is refined by recursively propagating from its neighbors:

$$e_v^l = f_{agg}\left(e_v^{l+1}, e_{N(v)}^{l+1}\right) \tag{2}$$

Finally, KG-based recommendation methods adopt a layer-aggregation $f_{concat}$ mechanism to concatenate the representations of each layer into a single vector:

$$e_v = f_{concat}(e_v^1, \ldots, e_v^l) \tag{3}$$

It is worth noting that vector $e_v$ is simply a weighted summation of entities linked to node $v$ and the key point is to obtain a function $f$ learned by collaborative signals to output weight $w_i$:

$$e = \sum w_i e_i \tag{4}$$

Two factors are important in calculating $w_i$: First, preserving the features' integrity in the entity-relation chain route between the current entity and the target entity, differentiate each entity more when calculating weight $w_i$. Second, The grouping methods affect the ability to filter irrelevant entities. Grouping entities in appropriate ways enables the discriminator to aggregate better useful information. Typical group methods divide the entities linked to the same entity and place them in one group.

### 2.2. Limitations of existing models

Existing methods suffer from some limitations when calculating $w_i$, hindering the effective and efficient training of KG-based recommendation models.

- Limitation I: Existing methods cannot unify all entities and relations involved in the link between entity $i$ and entity $j$, including users and items, leading to feature leakage, e.g., KGAT implements a relational attention mechanism $\pi(h, r, t)$, ignoring the user and item information.
- Limitation II: Current attention mechanisms often follow the key-value paradigm and only perform two or three feature crossing, which is insufficient to calculate weight $w_i$.
- Limitation III: Existing approaches often make intra-group comparisons, which may not discover the effective entity for the recommendation task and filters the irrelevant entity.

# 3. Theory and framework

This section first introduces the terminology and describes the problem of recommender systems based on knowledge graphs. Then we present the proposed **D**ynamic **K**nowledge **S**elector and **E**valuator Network (**DKSE**), and then we apply the multi-stream attention mechanism to distill the effective information from the noise, which conducts feature crossing in multi-combinations. Finally, we describe the prediction layer and optimization algorithm formally. The architecture of DKSE is illustrated in Figure 2.

## 3.1. Task Definition

**User-item bipartite graph:** We regard user-item interaction data as a bipartite graph $G_b$. Specifically, given a user set $U = \{u_1, u_2 \dots\}$, item set $V = \{v_1, v_2 \dots\}$, and a user-item interaction matrix $Y = \{y_{u,v} | u \in U, v \in V \dots\}$, the items and users are denoted as nodes in the user-item bipartite graph and $y_{u,v} = 1$ indicates an edge connection between user $u$ and item $v$. Otherwise, there is no edge connection.

**Knowledge graph:** Formally, a knowledge graph, which contains real-world entities, is defined as a triplet $G_k = \{(h, r, t) | h, t \in E, r \in R\}$, where the head, relation, and tail entity are denoted as $h, r, t$, respectively. Besides, if we regard $e$ as any entity in the knowledge graph, then $N(e)$ is the set of neighbor entities of entity $e$ in $G_k$.

**Collaborative Unified graph:** The collaborative, unified graph $G_c$ combines the user-item bipartite graph and knowledge graph seamlessly through item-entity alignment.

**Chain Route:** Chain Route is the relation and entity involved in the link between entity $i$ and entity $j$ in a Collaborative Unified graph and is a directed sequence comprising entities and relations, such as $R = \{u_1, r_1, v_1, r_2, e_1\}$. Chain Route $R$ contains all information used to calculate the relevance between user $u_1$ and entity $e_1$.

**Task Definition:** We formulate the recommendation task as follows. Based on a Collaborative Unified graph $G_c$, we aim to learn the prediction function $f$ to predict the connection possibility between the user and the item.

## 3.2. Multi-Stream Attention Layer

The multi-stream attention layer(MSAL) comprises two major modules: **Knowledge Selector** and **Chain Route Evaluator**. The former explicitly chooses entities or relations (including use/item information) from the chain route that should be involved in generating the attentive weights. The latter learns from the collaborative signals how to suggest which chain route should be given more attention to capture knowledge associations more effectively. Each part is discussed exhaustively next.

*Knowledge Selector***:** Intuitively, surrounding entities of the user/item in a collaborative, unified graph can represent the user's preferences for an item. For example, the recommended movie Avengers: Endgame has six entities connected to it in the collaborative, unified graph, including director Anthony Russo, actor Robert Downey Jr, and actor Chris Evans, film Sherlock Holms, Iron Man, and Captain America. Each entity has a unique chain route containing all knowledge that could be beneficial for representing the user's preference for the movie Avenger: Endgame. In order to filter the less informative knowledge before evaluating all six chain routes for the recommendation, we initialize $n$ query vectors $q \in R^d$ to select useful entities and relations. Considering a chain route $C = \{u_1, r_1, v_1, r_2, e_1 \dots\}$, that is a collection of $n$ entities/relations, we apply the following nonlinear transformations $f_k$:

$$k = ReLU(w_k e + b_k) \tag{5}$$

where $w_k \in R^{d \times d}, b_k \in R^d$ are trainable weight matrices and $e \in R^d$ are initialized entity/relation/user/item vectors. Function $g$, e.g., inner product, is used to compute the score $\pi_e^q$:

$$\pi_e^q = g(q, k) \tag{6}$$

$$\widetilde{\pi_e^q} = \frac{\pi_e^q}{\sum_{e \in N(e)} \pi_e^q} \tag{7}$$

where $\widetilde{\pi_e^q}$ is the normalized score. We aggregate knowledge in the chain route with a normalized score, which acts as an automatic selector and generates selective features for each chain route $e_c$.

$$e_c = \sum_{e \in N(e)} \widetilde{\pi_e^q} e \tag{8}$$

**Chain Route Evaluator:** To effectively discriminate the contributions of different neighborhoods for the recommendation task, we use the chain route evaluator to obtain the neighborhood vector $e_{N(v)}$. Considering entity $v$ in the Collaborative Unified graph with $m$ chain routes $C_v = \{C_1, C_2, C_3 \dots C_m\}$, we compute $\widetilde{s_c}$ to characterize the importance of the neighborhood entities:

$$s_c = w_c e_c + b_c \tag{9}$$

$$\widetilde{s_c} = \frac{s_c}{\sum_{e \in C_v'} s_c} \tag{10}$$

$$e_N = \sum_{e \in N} \widetilde{s_c} e \tag{11}$$

where $\widetilde{s_c}$ is the normalized score and $C_v'$ is a subset of $C_v$. We explore several methods to divide entities into groups during normalization, which affects how entities are compared. Currently, some typical grouping methods exist, e.g., dividing all entities into one group and normalizing the score. Thus, we design four methods to find the appropriate one.

- Global Grouping. Map all entities into one group.
- Vertical Grouping. Divide all entities in one chain route into a group.
- Horizontal Grouping. Divide all entities in the same depth from $v$ into a group.

## 3.3. Prediction

After applying the knowledge selector and chain route evaluation, we obtain the user/item vector $\widetilde{e_u}/\widetilde{e_v}$ by adding a neighborhood vector $e_{N(v)}$:

$$\widetilde{e_u} = e_u + e_{N(u)} \tag{12}$$

$$\widetilde{e_v} = e_v + e_{N(v)} \tag{13}$$

We calculate the predicted click probability $\widehat{y_{u,v}}$ with $\widetilde{e_u}/\widetilde{e_v}$:

$$\widehat{y_{u,v}} = \sigma(u^T v) \tag{14}$$

where $\sigma$ is the sigmoid function. The entire algorithm process is presented below:

| Algorithm: DKSE algorithm |
|---|

**Input**: Interaction matrix $Y$; knowledge graph $G_k$;
trainable parameters: $\{u\} \in U, \{e\} \in E, \{r\} \in R,$
$Q = \{q_1 \dots q_n\}, \{w_c, b_c\}, \{w_k b_k\};$
hyper-parameters: $l_u, l_v, n_u, n_v, n, d, \lambda;$
**Output**: Prediction function $F(u, v | \Theta, Y, G_k)$
**1:** while DKSE not converge do
**2:**   for $(u, v)$ in $Y$ do
**3:**     $e_{N(u)} \leftarrow \pmb{MSAL(u, N(u))}$
**4:**     $e_{N(v)} \leftarrow \pmb{MSAL(v, N(v))}$
**5:**     $\widetilde{e_u} \leftarrow e_u + e_{N(u)}$
**6:**     $\widetilde{e_v} \leftarrow e_v + e_{N(v)}$
**7:**     **Calculate predicted click probability** $\widehat{y_{u,v}}$
**8:**     **Update parameters by gradient descent.**
**9:** ***Return*** ***F***

**10:** **Function** ***MSAL(e, N(e))***
**11:**   for $q_i$ in $Q$ do
**12:**     $e_c \leftarrow Knowledge\ Selector(e, N(e))$
**13:**     $\widetilde{s_{c_i}} \leftarrow Chain\ Route\ Evaluator(e_c, N(e))$
**14:**     **calculate the mean value** $\widetilde{s_c}$ of $\{\widetilde{s_{c_1}} \dots \widetilde{s_{c_n}}\}$
**15:**     $e_{N(e)} \leftarrow \widetilde{s_c} \cdot N(e)$
**16:** **Return** $e_{N(e)}$

### 3.4. Loss function

DKSE's loss function comprises two parts. First, we calculate the cross-entropy loss $\mathcal{L}_{base}$ between the user click probability and label, defined as follows:

$$\mathcal{L}_{base} = \sum_{u \in U} \left( \sum_{v \in \{v | (u,v) \in D\}} \Phi(\widehat{y_{u,v}}, y_{u,v}) \right) \quad (15)$$

where $\Phi$ is the cross-entropy loss, and $D$ denotes the training dataset discussed in detail in the Dataset Description Section. Second, we randomly select a user-item pair without interaction in a batch data $B$ as negative samples and calculate the contrastive loss $\mathcal{L}_{cl}$:

$$\mathcal{L}_{cl} = -\log \frac{e^{\frac{\sigma(u^T v)}{\tau}}}{\sum_{u \in U} \sum_{v \in \{v | (u,v)D\}} e^{\frac{\sigma(u^T v)}{\tau}}} \quad (16)$$

Moreover, we add an L2 regularization term $\mathcal{F}$ for all parameters to alleviate the model from overfitting. Finally, we use the Adam optimizer to optimize all parameters.

$$\mathcal{L} = \mathcal{L}_{base} + \mathcal{L}_{cl} + \lambda ||\mathcal{F}||_2^2 \quad (17)$$

### 3.5. Time Complexity Analysis

The computational complexity of DKSE is divided into two parts: KGIR and UER. Given the item sampling depth $l_k$ for a neighborhood sample size $M$, and the user sampling depth $l_p$ for the neighborhood sample size $N$ and the dimension of the embedding size $d$, the computational complexity for KGIR and UER are $O(l_k Md)$ and $O(l_p Nd)$, respectively. Thus, the overall training complexity of DKSE is $O(l_k Md + l_p Nd)$.

We conducted several experiments on an RTX-3060 GPU and compared the training speed of DKSE against the baseline models KGAT [29], KGCN [30], and RippleNet [32]. The training time per epoch for

DKSE, KGAT, KGCN, RippleNet, and BPRMF on Amazon-book is about 20s, 600s, 1.9s, 2.7s, and 4s, respectively.

## 4. Experiments

This section evaluates the performance of DKSE on three real-world datasets and verifies its effectiveness. Moreover, we aim to answer the following questions:

- RQ1: How does DKSE perform compared to the state-of-the-art baseline models?
- RQ2: How do different components affect DKSE's performance?
- RQ3: How do different parameter settings affect DKSE?

### 4.1. Dataset Description

To evaluate the effectiveness of DKSE, we utilize three publicly accessible real-world datasets, MovieLens-1M, LFM-1b, and Amazon-book, which are varied in scenarios, size, and sparsity. The statistics of the three datasets are listed in Table 1, and descriptions of the datasets are as follows:

**Table 1 Dataset Statistics**

| | | *ML-1M* | *LFM-1b* | *AZ-book* |
|---|---|---|---|---|
| **User-item Interaction Network** | #Users | 6036 | 12134 | 6969 |
| | #Items | 2445 | 15471 | 9854 |
| | #Interactions | 753772 | 2979267 | 552706 |
| | #Avg user clicks | 124.9 | 152.3 | 79.3 |
| | #Avg clicked items | 308.3 | 119.4 | 56.1 |
| **Knowledge Graph** | #Source | Microsoft Satori | Freebase | Freebase |
| | #Entities | 182011 | 106389 | 113487 |
| | #Relations | 12 | 9 | 39 |
| | #Triples | 1241995 | 464567 | 2557746 |

- **MovieLens-1M**: A widely used benchmark dataset for movie scenarios, containing about 1 million explicit ratings (from 1 to 5) for 2445 items by 6036 users.
- **LFM-1b**: A dataset collected from the online music system Last.fm that records user listening history, e.g., artists and albums, containing about 3 million detailed rating records for 15,471 songs by 12,134 users.
- **Amazon-book**: It contains information such as users, items, ratings, and rating time and has about 500,000 rating records for 9,854 books by 7,000 users, widely used for book recommendation.

We transform the explicit interactions into implicit feedback among the above datasets. Specifically, we indicate positive/negative samples as 1/0 instead of using raw ratings as labels. For MovieLens-1M, ratings over 4 are considered positive, and below 4 as negative. For the LFM-1b and Amazon-book datasets, all rating records are considered positive samples, and we randomly sample from unwatched interactions to obtain the negative data. To ensure the quality of the datasets, we applied a 20-core setting, i.e., discard users and items with less than 20 interactions. We randomly select 60%, 20%, and 20% of historical interactions for each dataset to constitute the training/ validation/test datasets.

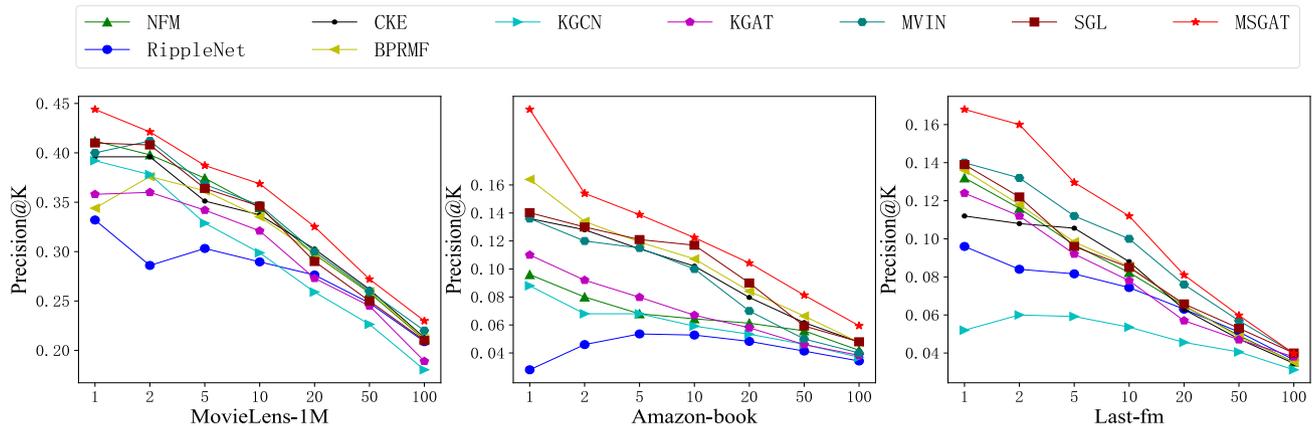

**Figure 2. Precision@K, NDCG@K in top-K recommendation on all datasets, and all experiments are performed 3 times.**

## 4.2. Baselines

DKSE is challenged against the following baselines:

- **KGCN** [30]: An end-to-end framework that utilizes graph convolution networks to encode high-order structural and semantic entity information from knowledge graphs to enhance item representations.
- **RippleNet** [32]: This method is similar to a storage network, which represents a user by his click history through integrating knowledge graphs into recommender systems, overcoming the limitations of existing embedding-based and path-based KG-aware recommendation methods.
- **KGAT** [29]: A recommendation model based on a graph neural network, which uses a hybrid structure of knowledge graph and user-item graph as a collaborative knowledge graph to extract higher-order connections in CKG in an end-to-end manner. KGAT uses an attention mechanism to distinguish the importance of neighbors.
- **NFM** [11]: A factorization-based method that employs the Bilinear Interaction pooling operation to learn low-level combination features.
- **BPRMF** [12]: A pairwise sorting algorithm based on matrix decomposition, which assumes that the parameters obey a normal distribution to alleviate model overfitting.
- **CKE** [33]: A recommendation algorithm that combines knowledge graphs and multi-modal features into knowledge graphs, text descriptions, and cover images. In this paper, we only consider knowledge graphs as side information.
- **SGL** [34]: combines contrastive learning with a graph encoder to enhance the graph data by randomly removing and masking the features of nodes and edges to construct positive and negative samples so that the model learns more essential features.
- **MVIN** [35]: learns representations of items from multiple perspectives from both the user and entity perspectives, enriching the user-item interactions and refining the entity-entity interactions.

## 4.3. Experimental Settings

Our experimental parameter settings for DKSE are reported in Table 2:

**Table 2- Hyper-parameter Settings For DKSE**

| LFM-1b | $l_u = 2, n_u = 64, l_v = 1, n_v = 32, d = 64, \lambda = 1e\text{-}6, n = 6$ |
|---|---|
| MovieLens-1M | $l_u = 1, n_u = 32, l_v = 2, n_v = 32, d = 32, \lambda = 1e\text{-}5, n = 4$ |
| Amazon-book | $l_u = 2, n_u = 8, l_v = 3, n_v = 32, d = 64, \lambda = 1e\text{-}5, n = 4$ |

The values of the L2 regularization coefficient λ and the vector dimension d range from [1×10-3, 1×10-4, 1×10-5, 1×10-6, 5×10-6,1×10-7] and [4, 8, 16, 32,64], respectively. Moreover, we set the number of neighbor samples to 16, 8, and 64 for KGAT, KGCN, and RippleNet, the number of hops to 2 for KGAT and RippleNet on all datasets, and [2,2,1] for KGCN on MovieLens-1M, Amazon-book, and LFM-1b. The hardware setup for all models is an i7-10700 processor and an RTX3060 GPU.

## 4.4. Performance Comparison (RQ1)

We experimentally compare all baseline models (CKE, NFM, BPRMF, RippleNet, KGCN, KGAT, SGL, MVIN) with DKSE on three datasets, using the AUC, ACC, and F1 performance metrics, widely used in CTR prediction. The corresponding experimental results are reported in Table 3. We also use the Precision and NDCG as the evaluation indicators for the TOP-K recommendation task, with the experimental results illustrated in Figure 3. Intuitively, our proposed DKSE outperforms all state-of-the-art competitor models on all datasets. From the experimental results, the following can be observed:

- According to Table 3, DKSE achieves a significant performance improvement on all datasets. Specifically, compared with the best baseline, the AUC of DKSE on MovieLens-1M, music, and Amazon-book is increased by 0.54%, 2.27%, and 3.95%, respectively.
- The DKSE model has the greatest improvement on the Amazon-book dataset and achieved the lowest performance on the MovieLens-1M dataset due to the relatively high average clicked items compared with other datasets. Moreover, our model's performance is heavily correlated with the average clicked items of the dataset. When the average clicked items are higher, there is less improvement space for DKSE because for higher average clicked items, the users are more likely to click the same items. Thus, when we use the user click history to express the interests and preferences of different users, it is more difficult to distinguish them.

**TABLE 3. AUC, ACC, AND F1 RESULTS IN CTR PREDICTION**

| Models | LFM-1b | | | MovieLens-1M | | | Amazon-book | | |
|---|---|---|---|---|---|---|---|---|---|
| | AUC | ACC | F1 | AUC | ACC | F1 | AUC | ACC | F1 |
| NFM | 0.946(-2.56%) | 0.879(-3.91%) | 0.916(-2.04%) | 0.925(-1.03%) | 0.852(-1.01%) | 0.854(-1.45%) | 0.842(-7.67%) | 0.764(-7.99%) | 0.770(-8.18%) |
| CKE | 0.948(-2.36%) | 0.881(-3.69%) | 0.917(-1.94%) | 0.926(-0.92%) | 0.852(-1.01%) | 0.854(-1.45%) | 0.836(-8.32%) | 0.745(-10.27%) | 0.767(-8.54%) |
| BPRMF | 0.946(-2.56%) | 0.878(-4.02%) | 0.915(-2.51%) | 0.921(-1.46%) | 0.850(-1.24%) | 0.851(-1.80%) | 0.832(-8.76%) | 0.742(-10.63%) | 0.764(-8.90%) |
| RippleNet | 0.938(-3.39%) | 0.887(-3.04%) | 0.920(-1.61%) | 0.920(-1.56%) | 0.852(-1.01%) | 0.850(-1.92%) | 0.820(-10.08%) | 0.745(-10.27%) | 0.750(-10.5%) |
| KGCN | 0.917(-5.55%) | 0.865(-5.44%) | 0.906(-3.11%) | 0.907(-2.95%) | 0.832(-3.33%) | 0.836(-3.53%) | 0.808(-11.39%) | 0.733(-11.72%) | 0.741(-11.6%) |
| KGAT | 0.937(-3.49%) | 0.875(-4.35%) | 0.911(-2.58%) | 0.921(-1.46%) | 0.846(-1.71%) | 0.848(-2.15%) | 0.852(-6.57%) | 0.780(-6.06%) | 0.798(-4.84%) |
| SGL | 0.948 (-2.36%) | 0.878 (-4.02%) | 0.911 (-2.58%) | 0.930 (-0.49%) | 0.8541(-0.7%) | 0.8545(-1.4%) | 0.891 (-2.29%) | 0.799(-3.77%) | 0.797(-4.96%) |
| MVIN | 0.965 (-0.61%) | 0.910 (-0.52%) | 0.932(-0.33%) | 0.9318(-0.3%) | 0.8573(0.40%) | 0.8572(1.08%) | 0.875(-4.05%) | 0.793 (-4.49%) | 0.793(-5.44%) |
| **DKSE** | 0.9709 | 0.9148 | 0.9351 | **0.9346** | **0.8607** | **0.8666** | 0.9119 | 0.8303 | 0.8386 |

**TABLE 4 DKSE ablation study results. We conduct experiments on eight components of modules(UEI, IUPP) on three datasets and present the effect of different components. These components are U/R、U/H in UEI and U/H、U/R、U/T、V/T、V/H in IUPP.**

| Ablation component | Datasets | | | | | | |
|---|---|---|---|---|---|---|---|
| | U/V | H | R | T | LFM-1b | AZ-book | ML-1M |
| N/A | | | | | 0.9709 | 0.8354 | 0.881 |
| w/o U/V | ✔ | | | | 0.9123 | 0.8990 | 0.871 |
| w/o H | | ✔ | | | 0.9346 | 0.8732 | 0.930 |
| w/o R | | | ✔ | | 0.940 | 0.884 | 0.921 |
| w/o T | | | | ✔ | 0.9587 | 0.89 | 0.912 |

- Figure 3 highlights that in the top-K recommendation task, DKSE has achieved the best performance in Precision@K on LFM-1b, followed by Amazon-book and MovieLens-1M. When K is 1, 2, or 5, the performance of DKSE on the LFM-1b is significantly improved compared with the other baseline models. As K increases, this improvement shows a slightly decreasing trend. For k>1, the DKSE has no distinct advantage over the competitor models on MovieLens-1M. To compensate for the insensitivity of Precision@K regarding the order of the sorting positions, we record the NDCG@K of all experiments. On all datasets, DKSE outperforms the baseline models in NDCG@K. To our surprise, although DKSE does not perform as well as expected in Precision@K on MovieLens-1M, it still presents an appealing performance in NDCG@K. This is because DKSE is still in the lead in AUC compared to the baseline models, which often reflects the model's sorting ability.

- Table 3 suggests that the performance of all baseline models on the three datasets are ranked from high to low as follows: LFM-1b, MovieLens-1M, and Amazon-book, potentially due to the average interaction of each user in the three datasets. The LFM-1b has the highest average user clicks, and the higher they are, the stronger the model's generalization ability. Regarding LFM-1b and MovieLens-1M,

- the best performance of the baseline models is ranked as CKE, NFM, BPRMF, then RippleNet, and KGAT, and the worst is KGCN. On the Amazon-book, the ranked performance of the baseline models is KGAT, NFM, CKE, BPRMF, RippleNet, and KGCN.

- On the LFM-1b and MovieLens-1M, the CF-based method is better than the KG-based methods. NFM utilizes a DNN module based on FM, which extracts deep cross features through a nonlinear transaction. BPRMF based on MF aims to model the user's relative preference over different items. In particular, CKE, which fuses CF and KG, performs particularly well on the above two datasets because KG and CF combination can enhance the representation of items. At the same time, we observed that the performance of CKE has declined on the Amazon-book dataset, which has richer KG entity connections. This is potentially due to the limitation of the KG representation learning methods, which cannot handle complex graphs well.

- By comparing the KG-based models CKE, RippleNet, KGAT, and KGCN, we discover that KGCN performs the worst on three datasets because it only considers the neighborhood entities of the items as input in the knowledge graph and does not integrate the user-item bipartite graph with item KG. Moreover, RippleNet outperforms KGAT on the LFM-1b dataset because RippleNet mainly uses the user click history to enhance the user representation, which is more suitable for the LFM-1b with the largest number of user average clicks.

### 4.5. Study of DKSE (RQ2)

Next, we evaluate the effectiveness of different group methods of DKSE by conducting some experiments. We experiment with the four model variants — DKSE(Hor), DKSE(Glo), DKSE(Ver), and DKSE(base), to analyze the impact of the group method. Next, to further explore the influence of different sub-components, we split DKSE into 5 sub-components for detailed comparison—w/o U/V, w/o H, w/o R, w/o T, and w/o CL.

**Table 5 -  AUC of DKSE w.r.t different Group Method.**

|         | DKSE(Gol) | DKSE(Ver) | DKSE(Hor) |
|---------|-----------|-----------|-----------|
| AZ-book | **0.9123**  | 0.8940    | 0.9119    |
| ML-1M   | **0.9346**  | 0.9177    | 0.9341    |
| LFM-1b  | **0.9709**  | 0.9409    | 0.9701    |

- **Group Method Comparison**
  Here, we design three model variants, i.e., DKSE(Hor), DKSE(Glo), DKSE(Ver), DKSE(base) to perform experiments, which provide the following observations: Figure 4 highlights that on three datasets, the rank of AUC is DKSE(Hor), DKSE(Glo), DKSE(Ver), and DKSE(base). The BASE model, which discards comparison between entities, has the lowest AUC. Solely DKSE(Hor) positively affects the results, while DKSE(Glo) achieves the best performance as it exploits all entities comparison.

- **Comparison subcomponents**
  We conduct experiments on five components of three datasets, with the experimental results reported in Table 4. The following conclusions can be drawn:

  **w/o U/V information:** On LFM-1b, **U/V** contributes the most, while LFM-1b has the highest user average click number of 152.3. IUPP(V/T) compares the similarity between the item and the user's historically clicked items, which performs better when the user-clicked item sequences are longer.

  **w/o R information:** On the Amazon-book, this component makes the most contributions compared with the other components because this database has the most categories of relations, which is beneficial to aggregate the neighborhood entities of the item. In other words, this component helps aggregate entity information in a finer-grained and personalized way by meticulously depicting user preferences for different relations.

  **w/o T information:** On the MovieLens-1M, this component is the most prominent. Compared with the other two datasets, KG in MovieLens-1M has the largest number of entities when aggregating entity information. The more entities, the more the T component contributes to discovering user perception of different entities.

### 4.6. Parameter Sensitivity (RQ3)

Next, we investigate the effects of different parameters on DKSE.

**Table 6 - AUC of DKSE w.r.t different number of $n_u, l_u$.**

|         | $n_u \backslash l_u$ | 1 | 2 | 3 | 4 |
|---------|------|-------|-------|-------|-------|
| AZ-book | 4    | .9025 | .9042 | .9048 | .9055 |
|         | 8    | .9050 | **.9096** | .9068 | .9073 |
|         | 16   | .9054 | .9050 | .9035 | .9053 |
|         | 32   | .9059 | .9042 | .9003 | .9023 |
|         | 64   | .9057 | .9047 | .8997 | .9025 |
| ML-1M   | 4    | .9322 | .9326 | **.9322** | .9320 |
|         | 8    | .9325 | .9331 | .9322 | .9323 |
|         | 16   | .9329 | .9328 | .9322 | .9327 |
|         | 32   | **.9332** | .9320 | .9306 | .9317 |
|         | 64   | .9329 | .9322 | .9296 | .9310 |
| LFM-1b  | 4    | .9675 | .9676 | .9679 | .9678 |
|         | 8    | .9681 | .9683 | .9682 | .9685 |
|         | 16   | .9686 | .9690 | .9692 | .9689 |
|         | 32   | .9688 | .9695 | .9685 | .9693 |
|         | 64   | .9691 | **.9696** | .9686 | .9690 |

**Table 7- AUC of DKSE w.r.t different sampling size $n_v, l_v$.**

|         | $n_v \backslash l_v$ | 1 | 2 | 3 | 4 |
|---------|------|-------|-------|-------|-------|
| AZ-book | 4    | .8991 | .8999 | .8996 | .8988 |
|         | 8    | .9068 | .9067 | .9072 | .9060 |
|         | 16   | .9101 | .9098 | .9103 | .9100 |
|         | 32   | .9101 | .9088 | **.9104** | .9010 |
|         | 64   | .9096 | .9094 | **.9104** | .9099 |
| ML-1M   | 4    | .9326 | .9332 | .9332 | .9330 |
|         | 8    | .9331 | .9336 | .9341 | .9340 |
|         | 16   | .9337 | .9342 | .9344 | .9343 |
|         | 32   | .9342 | **.9345** | **.9345** | .9342 |
|         | 64   | .9340 | **.9345** | **.9345** | .9348 |
| LFM-1b  | 4    | .9674 | .9671 | .9674 | .9676 |
|         | 8    | .9696 | .9692 | .9693 | .9693 |
|         | 16   | .9702 | .9697 | .9696 | .9694 |
|         | 32   | **.9703** | .9693 | .9696 | .9694 |
|         | 64   | .9701 | .9687 | .9693 | .9690 |

**Table 8 AUC of DKSE w.r.t $n$ query vectors.**

| $n$     | 1     | 2     | 4     | 6     | 8     |
|---------|-------|-------|-------|-------|-------|
| AZ-book | .9103 | .9119 | **.9132** | .9125 | .9123 |
| ML-1M   | .9345 | .9345 | **.9351** | .9348 | .9346 |
| LFM-1b  | .9699 | .9703 | .9708 | **.9709** | .9707 |

- **Sampling depth.** Table 5 reports the performance of DKSE under different user sampling depths $l_u$ and item sampling depths $l_v$. DKSE performs best at $l_u$ =2, 2, 1 and $l_v$ =1, 3, 2 on LFM-1b, Amazon-book, and MovieLens-1M, respectively.

- **Neighborhood sampling number.** Table 6 reports the performance of DKSE under different item neighborhood sampling numbers $n_v$ and user neighborhood sampling numbers $n_u$. DKSE performs best for $n_u$= 64, 8, 32 and $n_v$=32, 32, 32 on LFM-1b, Amazon-book, and MovieLens-1M, respectively.
- **Query vector number.** Table 7 presents the performance of DKSE under different query vector numbers. DKSE performs best when employing 6, 4, and 4 on LFM-1b, Amazon-book, and MovieLens-1M, respectively.

## 5. Conclusion and future work

This paper proposes a novel knowledge-distilling method called DKSE for recommendation, which comprises a Knowledge Selector and a Chain Route Evaluator. A knowledge Selector filters the less informative knowledge involved in the chain route, and the Chain Route Evaluator is used to select informative entities to represent user and item features for recommendation. Extensive experiments show the superiority of DKSE compared with several state-of-the-art models. In addition, we conduct a series of ablation experiments to demonstrate the effectiveness of each module.